\documentclass[runningheads,a4paper]{llncs}

\usepackage{amssymb}
\usepackage{graphicx}
\usepackage{helvet}
\usepackage{courier}
\usepackage{amsmath}
\usepackage{diagbox}
\usepackage{booktabs}
\usepackage{multirow}
\usepackage{subfig}
\usepackage{comment}
\usepackage{url}

\urldef{\mailsa}\path|{alfred.hofmann, ursula.barth, ingrid.haas, frank.holzwarth,|
\urldef{\mailsb}\path|anna.kramer, leonie.kunz, christine.reiss, nicole.sator,|
\urldef{\mailsc}\path|erika.siebert-cole, peter.strasser, lncs}@springer.com|    
\newcommand{\keywords}[1]{\par\addvspace\baselineskip
\noindent\keywordname\enspace\ignorespaces#1}

\begin{document}

\mainmatter  

\title{Can Online Emotions Predict \\ the Stock Market in China?}

\titlerunning{Can Online Emotions Predict the Stock Market in China?}

%
%
\author{Zhenkun Zhou\inst{1}
\and Jichang Zhao\inst{2*} \and Ke Xu\inst{1}}

\authorrunning{Can Online Emotions Predict the Stock Market in China?}

\institute{State Key Laboratory of Software Development Environment, Beihang University,\\
\and School of Economics and Management, Beihang University\\
\and * Corresponding author: jichang@buaa.edu.cn
}

%
%

\toctitle{Can Online Emotions Predict the Stock Market in China?}
\tocauthor{Authors' Instructions}
\maketitle

\begin{abstract}
Whether the online social media, like Twitter or its variant Weibo, can be a convincing proxy to predict the stock market has been debated for years, especially for China. However, as the traditional theory in behavioral finance states, the individual emotions can influence decision-making of investors, so it is reasonable to further explore this controversial topic from the perspective of online emotions, which is richly carried by massive tweets in social media. Surprisingly, through thorough study on over 10 million stock-relevant tweets from Weibo, both correlation analysis and causality test show that five attributes of the stock market in China can be competently predicted by various online emotions, like disgust, joy, sadness and fear. Specifically, the presented model significantly outperforms the baseline solutions on predicting five attributes of the stock market under the $K$-means discretization. We also employ this model in the scenario of realistic online application and its performance is further testified.   
\end{abstract}

\section{Introduction}

With explosive development of online social media, massive amounts of tweets are spreading in social media platforms like Twitter and Weibo. These tweets not only convey the factual information, but also reflect the emotions of the authors. Taking Weibo as an example, around 100 million Chinese tweets are posted every day and from which we can not only sense what happens in China, but also how 500 million users feel. In fact, the online social media indeed provide us an unprecedented opportunity to study the detailed human behavior from many views and the investment decision in the stock market is one of the most important issues, which already attracts much attention in recent decades. 

However, whether online social media like Twitter can be excellent predicator is still controversial, especially for the stock market in China~\cite{Bollen_predict_market,wanyun2013investors,mao2014quantifying}. Different from the west, the marking policy intervention in China will introduce more non-market factors that might disturb the fluctuation of the stock market. And moreover, those possible interventions could be leaked through the social media and then greatly influence the investors' emotions and choices. Besides, considering the irrationality of huge amount of individual investors in China, their actions might be more easily affected by online news and other investors' feelings about the market. Then the messages about the stock market and the sentiments they convey could be good indicators for the market prediction. Thus, like the conventional behavioral finance theory thinks, which the emotion can influence the decision-process of the investors, it is necessary to investigate the following important issues:
\begin{itemize}
\item Is there indeed significant correlation between online emotions and attributes of Chinese stock market?
\item Can online emotions predict the attributes of the stock market in China?
\item Which emotion does play the critical role on predicting various attributes of the Chinese stock market? 
\end{itemize}

In the present study, we collect over 10 million Chinese stock-relevant tweets from Weibo and classify them into five emotions, including anger, disgust, joy, sadness and fear. Besides the daily closing index of Shanghai Stock Exchange\footnote{In the present work, index refers in particular to Shanghai Stock Exchange Composite Index and the trading volume refers in particular to the daily volume of the Shanghai Stock Exchange.}, we consider the daily opening index, the intra-day highest index, the intra-day lowest index and the daily trading volume of the stock market. By the correlation analysis and Granger causality test, it is revealed that disgust has a Granger causal relation with the closing index, joy, fear and disgust have Granger causal relations with the opening index, joy, sadness and disgust have Granger causal relations with the intra-day highest and lowest index,  and correlation between trading volume and sadness is unexpectedly strong. It's also surprising to find that anger in online social media possesses the weakest correlation or even is no relation with the Chinese stock market.

Based on the findings, we develop classification-oriented predictors, in which different emotions are selected as features, to predict 5 daily attributes of the stock market in China. The comparison with other baseline methods show that our model can outperform them according to $K$-means discretization. And the model is also deployed in a realistic application and achieves the accuracy of 64.15\% for the intra-day highest index (3-categories) and the accuracy of 60.38\% for the trading volume (3-categories). Our explorations demonstrate that the online emotions, specially disgust, joy, sadness and fear, in Weibo indeed can predict the stock market in China.

\section{Related Work}

Emotion expression and stock fluctuation are usually boned together in the traditional theory and even in social media.

Behavioral economics studies the effects of social, emotional and psychological factors on the economic decisions of individuals and institutions and the consequences for market prices. It demonstrates that mood can affect individual behavior and decision-making of investors~\cite{dolan2002emotion,nofsinger2005social}. 

Owing to lack of effective measurement method of sentiment, stock prediction using emotions had been in dispute~\cite{NBERw13189,brown2004investor}. However, with the recent widespread presence of computers and Internet, public emotions can be extracted from data on online platforms. Using Twitter as a corpus, some researchers built sentiment classifiers, which are able to determine different sentiments for a tweet~\cite{twitter_various_techniques}~\cite{pak2010twitter}~\cite{baseline_method}. Specially on Sina Weibo platform, Zhao et al. trained a fast Naive Bayes classifier for Chinese emotion classification, which is now available online for temporal and spatial sentiment patter discovery~\cite{moodlens}.

In addition, there have long been controversies on predictive power of social media aiming at different fields~\cite{sakaki2010earthquake,gayo2012wanted}. In the field of finance, Bollen et al. found that public mood on Twitter could predict the Dow Jones Industrial Average\cite{Bollen_predict_market}. The public mood dimensions of Calm and Happiness seemed to have a predictive effect. However, the tweets they collected were associated with whole social status, not just the stock market in America, which could not represent online investors' sentiment. Oh et al. also showed stock micro-blog sentiments did have predictive power for market-adjusted returns. Instead of emotion on social media~\cite{oh2011investigating}, some researchers  examined textual representations in financial news articles for stock prediction~\cite{Schumaker2009bu}~\cite{cohen2013mood}. Ding et al. proposed a deep learning method for event-driven stock market prediction on large-scale financial news dataset~\cite{XiaoDing:2015uo}. Besides, Bordino et al. showed that daily trading volumes of stocks traded in NASDAQ100 were correlated with daily volumes of queries related to the same stocks~\cite{bordino2012web}. 

However, to the best of our knowledge, related work referring to the stock market in China are relatively few. Mao et al. pointed out that Twitter did not have a predictive effect, as regards predicting developments in Chinese stock markets~\cite{mao2014quantifying}. They advised adopting the tweets on Weibo platform to research Chinese stock market. Based on 66,317 tweets of Weibo with one year and two emotion categories, Cheng and Lin found that the investors' bullish sentiment of social media can help to predict the stock market trading volume, but not stock market returns~\cite{wanyun2013investors}. Because of less collected data set and simple emotion classification, it is not easy to generalize their conclusion to other scenarios.

While in this paper, we focus purely on the stock market in China and try to understand the predictive ability of multiple online emotions in Weibo. Different from the previous study, we hope to develop predictors from more data sets and more sentiments and to predict more attributes of the real market. 

\section{Data sets}

\subsection{Online stock market emotions}
The feelings of investors can be collected through many different approaches, like questionnaire survey in previous study. While with the explosive development of the social media in China, more and more investors express their seeings, hearings and feelings on Weibo. Therefore, we choose and utilize the characteristic of Weibo to obtain online emotion referring to Chinese stock market.

From December 1st 2014 to December 7th 2015, the massive public tweets on Weibo are collected through its open APIs. However, only a fraction of the tweets are semantically related with Chinese stock market. Filtering out the irrelevant tweets and remaining the data that truly represent the stock market emotion is a very significant step. Therefore, we manually select six Chinese keywords, including Stock, Stock Market, Security, The Shenzhen Composite Index, The Shanghai Composite Index and Component Index with help of expertise from the background of finance. These important keywords are viewed that can depict the overall status of Chinese stock market. We postulate that if one tweet contain one or more of the six keywords selected, it likely describe the news, opinion or sentiment about Chinese stock market. In our database, the number of tweets related to stock market, involving one or more keywords, is a total of 10,550,525 from December 1st 2014 to December 7th 2015.

\begin{figure}
\centering
\includegraphics[height=5cm]{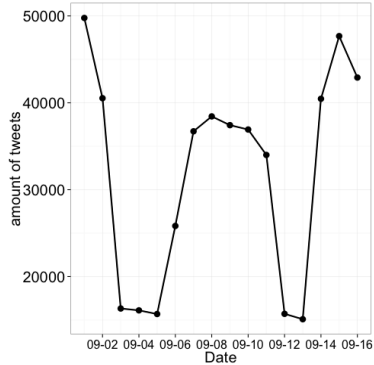}
\caption{Volume of the tweets related to the stock market from September 1st to 16th in 2015. There are respectively Memorial Day day between 9-3 and 9-5 and a weekend between 9-12 and 9-13, which are also non-trading days.}
\label{fig:amount_of_tweets}
\end{figure}

In the present study, the emotions are divided into five categories, including anger, sadness, joy, disgust and fear. In our previous work~\cite{moodlens}, a fast Naive Bayes classifier is trained on Weibo data for emotion classification. The system named MoodLens whose vital part is the emotion classifier, is now available online for temporal and spatial sentiment patter discovery. We arrange the tweets related to stock market, with one day as the time unit, and employ the system to label them with the emotions. There are five online stock market emotion time series: $X_{anger}, X_{sadness}, X_{joy}, X_{disgust}$ and $X_{fear}$. Online emotion is represented by $X$, which can be written as  

\begin{align*}
\small
 X = (X_{anger}, X_{sadness}, X_{joy}, X_{disgust}, X_{fear}).
\end{align*}

Observing the time series, the volume of tweets reduces significantly on non-trading days. Figure~\ref{fig:amount_of_tweets} shows the volume of the tweets related to the stock market from September 1st to 16th in 2015. There are separately Memorial Day between September 3rd and 5th and a weekend between September 12th and 13th, which are also non-trading days. We consider that online stock market emotion on non-trading days could not help us analyze and predict Chinese stock market. Hence, removing the data items on non-trading days from the time series, the results retain significant emotion data. It also partly reflects that tweets selected by the keywords could represent the stock market.

\begin{figure}
\centering
\includegraphics[width=7.5cm]{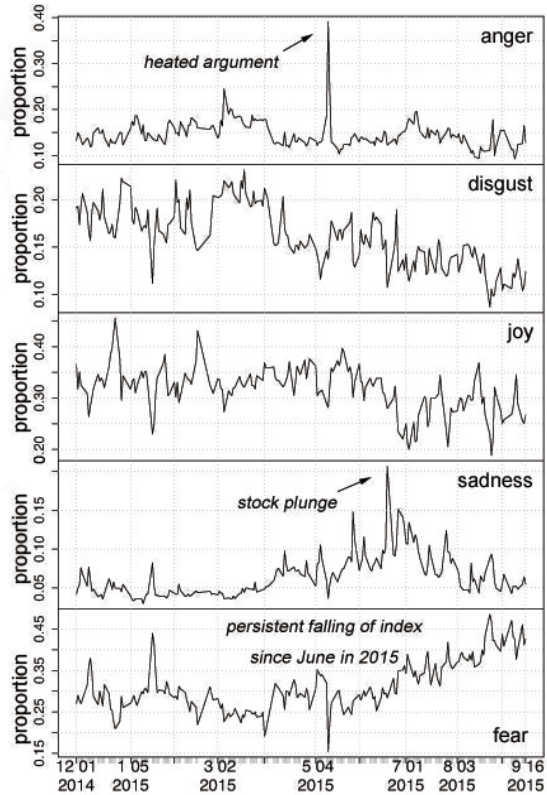}
\caption{Time series of each emotion from December 1st 2014 to September 16th 2015.}
\label{fig:emotion}
\end{figure}

For the sake of stability of online data, we measure the relative value (proportion) of each mood on one day as the final online stock market emotion $X$. Figure~\ref{fig:emotion} shows online stock market emotion time series $X$ from September 1st to 16th in 2015. We observe the spike in $X_{anger}$ on May 12th in 2015, when there is a heated argument between CEOs of listed companies. On June 19th in 2015, there was a plunge in Chinese stock market, with a fall of the index with 6.41\% and $X_{sadness}$ arrived the maximum. Since June in 2015, persistent falling of the index caused inward fears of investors, which can be seen from the sharp growth of $X_{fear}$ in Figure~\ref{fig:emotion}.

From the above observations, it can be concluded that the fluctuation of the sentiments can be connected with remarkable events in the stock market. It further inspires us to investigate the correlation and even causality between emotions and the market, which could provide the foundation for the predicting models.

\subsection{Stock market data}
In China, the economists and traders regard the Shanghai Stock Exchange Composite Index as depicting the overall status of the Chinese stock market. Therefore, the index is selected as price attribute of the stock market to analyze and predict. In particular, there are four values in candlestick charts of the index, which are respectively the closing index, the opening index, the intra-day highest index, the intra-day lowest index. We transform the values of the index into $Close$, $Open$, $High$ and $Low$ (to express rate of change on $i$-th day), and they can be written as
\begin{gather}
\small
\begin{split}
 Close_{i} = \frac{Index_{close,i}-Index_{close,i-1}}{Index_{close,i}} \times 100, \\
 Open_{i} = \frac{Index_{open,i}-Index_{close,i-1}}{Index_{close,i}} \times 100, \\
 High_{i} = \frac{Index_{high,i}-Index_{close,i-1}}{Index_{close,i}} \times 100, \\
 Low_{i} = \frac{Index_{low,i}-Index_{close,i-1}}{Index_{close,i}} \times 100.
\end{split}
\end{gather}
In addition to these four attributes, the trading volume of Shanghai Stock Exchange is also a key target used to reflects the status of the Chinese stock market. The time series of trading volume on each day is not transformed at all. 

We crawl historical data of the index and trading volume from December 1st 2014 to December 7th 2015. In this period, the number of trading days is totally 249 in our research. As a result, we obtain five time series which depict stock market's state on each day including $Y_{close}, Y_{open}, Y_{high}, Y_{low}$ and $Y_{volume}$. Each time series is a column vector of $Y$ (shown in Figure~\ref{fig:discretiztion}), i.e.,
\begin{align*}
\small
 Y = (Y_{close}, Y_{open}, Y_{high}, Y_{low}, Y_{volume}).
\end{align*}

The dataset ($X$ and $Y$) is devided into two parts according to the date: the 80\% data for training (from December 1st 2014 to September 16th 2015) and the 20\% data for testing (from September 17th to December 7th in 2015). The train set is used to not only analyze the relation between online emotions and the stock market but also fit and estimate the prediction model. The test set is kept in a vault and brought out only at the end of evaluation in realistic application.

\section{Correlation between online emotions and the stock market}
The preceding part of the paper describes the two groups of time series (in the train set): $X$ (represents online emotions, refers in particular to online stock market emotions) and $Y$ (represents the stock market), which contribute to discuss the correlation between online emotions and the stock market. However, the purpose of the paper is to find out whether online emotions can predict the stock market in China. Supposing that online emotions ahead of 1 to 5 days are available for stock prediction, we shifted emotion series to an earlier date: 1 to 5 days. Hence, each emotion corresponds to 5 time series according to shifted time. Each category of online emotions can be defined as (the categories of emotions are represented by $e$, $e=anger, sadness, joy, disgust$, or $fear$)
\begin{align*}
\small
 X_{e} = (X_{e, 1}, X_{e, 2}, X_{e, 3}, X_{e, 4}, X_{e, 5}).
\end{align*}

For the analysis the relation of $X$ and $Y$ ($T$ represents one certain time series of $X$ or $Y$), we normalize all the time series, of which data items are transformed to the values from 0 to 1 as
\begin{equation}
\small
 T_{i} = \frac{T_{i} - T_{min}}{T_{max} - T_{min}},
\end{equation}
$T_{i}$ is the $i$-th item in time series $T$, $T_{max}$ is the maximal value of $T$, and $T_{min}$ is the minimal value of $T$. Then, by using Pearson correlation analysis, we measure the linear dependence between $x$ ($X_{e, t}$, the emotion $e$ ahead of $t$ days in $X$) and $y$ (one target of $Y$) as Eq. (\ref{pearson}). $\rho$ is the Pearson correlation coefficient of time series $x$ and $y$ defined as
\begin{equation}
\small
 \rho = \frac{\Sigma(x_{i} - \bar{x})(y_{i} - \bar{y})}{\sqrt{\Sigma(x_i - \bar{x})^2\Sigma(y_i - \bar{y})^2}}.
\label{pearson}
\end{equation}

\begin{figure}
\centering
\subfloat[$Y_{close}$ and $X$]{\includegraphics[width=0.28\textwidth]{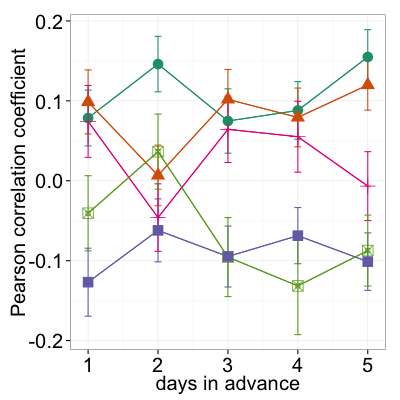}
\label{}}
\hfil
\subfloat[$Y_{open}$ and $X$]{\includegraphics[width=0.28\textwidth]{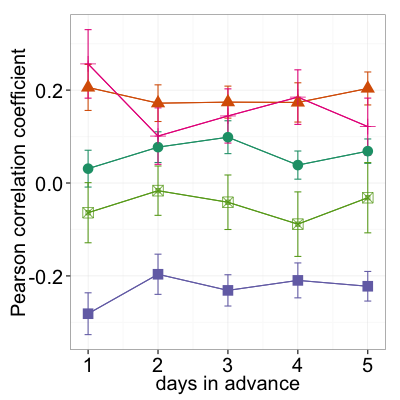}
\label{}}
\hfil
\subfloat[$Y_{high}$ and $X$]{\includegraphics[width=0.28\textwidth]{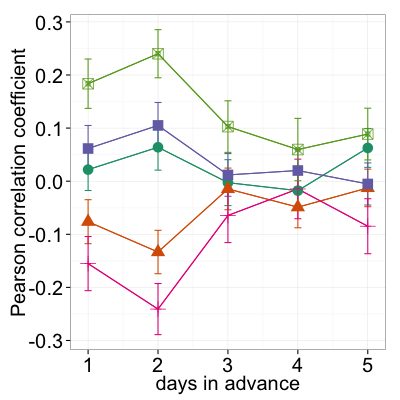}
\label{}}
\hfil
\subfloat[$Y_{low}$ and $X$]{\includegraphics[width=0.28\textwidth]{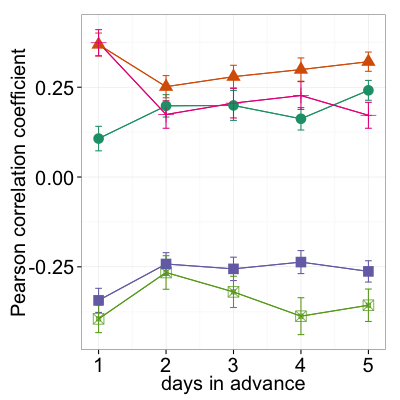}
\label{}}
\hfil
\subfloat[$Y_{volume}$ and $X$]{\includegraphics[width=0.36\textwidth]{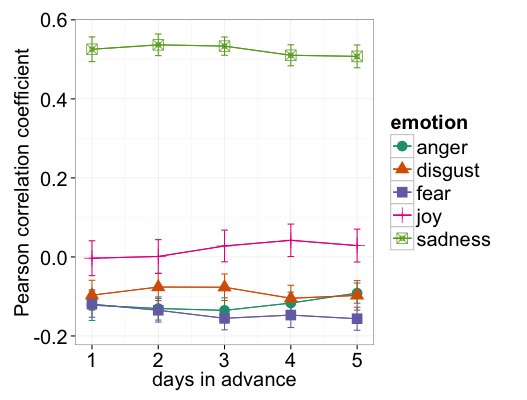}
\label{}}
\caption{Pearson correlation coefficient between five targets of stock market and online emotion time series (mean values of the coefficient in sampling results).}
\label{fig:error_bar}
\end{figure}

For observing whether there are distinct differences of correlation coefficient between online emotions and the stock market, the emotion time series associated with stock market time series are sampled 100 times. In one time, we sample randomly 150 pairs (from 191 pairs in the train set) of data items respectively from emotion time series and stock market time series. We calculate 100 sampling results' correlation coefficient, and then obtain the mean values and standard deviations. Figure~\ref{fig:error_bar} shows the means and error bars, which depicts sampling results' correlation coefficient. It can be seen that there are significant differences between different emotions.

In addition, we randomly shuffle the time series 100 times and calculate the Pearson correlation coefficient of them. Comparing the mean coefficients with that of non-shuffled time series, we find that all the correlation coefficients (shuffled) are near 0, and it suggests that most of correlation coefficients (not shuffled) are relatively higher than random results, indicating the significance of the correlation we find.

Through the correlation coefficients above (shown in Figure~\ref{fig:error_bar}), we set the threshold of correlation coefficient $\rho$ as 0.2 (the absolute value) and find some interesting and valuable results. The correlation between all online emotion time series (in $X$) and $Y_{close}$ is very low ($\rho<0.2$), which indicates little linear dependence between them. As to $Y_{open}$, the correlation coefficients with $X_{fear}$ (ahead of 1, 3, 4 and 5 days), $X_{joy}$ (ahead of 1 day) and $X_{disgust}$ (ahead of 1 and 5 days) are more than 0.2. $Y_{open}$ is negatively correlated with $X_{fear}$, positively correlated with $X_{joy}$ and $X_{disgust}$. As to $Y_{high}$, the correlation coefficients with $X_{joy}$ (ahead of 2 days) and $X_{sadness}$ (ahead of 2 days) are more than the threshold. $Y_{high}$ is negatively correlated with $X_{joy}$, positively correlated with $X_{sadness}$. $Y_{low}$ and 5 types of emotion time series have relatively high correlation, and the correlation coefficients between $Y_{low}$ and $X_{sadness}$ (ahead of 1 and 4 days) is the highest ($|\rho|>0.4$). $Y_{low}$ is negatively correlated with $X_{anger}$, $X_{disgust}$ and $X_{joy}$, positively correlated with $X_{sad}$ and $X_{fear}$. An interesting finding is that the correlation between $Y_{volume}$ and $X_{sadness}$ (no matter ahead of how many days) is unexpectedly high, correlation coefficients $\rho$ of which is more than 0.5. Besides, $Y_{volume}$ and other online emotion time series don't have a comparatively strong ($\rho>0.2$) correlation.

\section{Granger causality test of online emotions and the stock market}
Despite the correlation analysis, we also preform the causality test further on the train data. Here we apply the econometric technique Granger causality test to study the relation between online emotions and the stock market. The Granger Causality Test is a statistical hypothesis test for determining whether one time series is useful in forecasting another. One time series $x$ is said to Granger-cause $y$ if it can be shown that $x$ provides statistically significant information about future values of $y$, usually through a series of $t$-tests and $F$-tests on lagged values of $x$. We perform the analysis according to models shown in Eq~\ref{eq:granger_1} and~\ref{eq:granger_2} for the period from December 1st 2014 to September 16th 2015.
\begin{equation}
\small
 y_{t} = \alpha + \sum_{i=1}^{n}\beta_{i}y_{t-i} + \epsilon_{t}, \\
\label{eq:granger_1}
\end{equation}
\begin{equation}
\small
 y_{t} = \alpha + \sum_{i=1}^{n} \beta_{i}y_{t-i} + \sum_{i=1}^{n} \gamma_{i}x_{t-i} + \epsilon_{t}. \\ 
\label{eq:granger_2}
\end{equation}

The Granger causality test could select only two time series as inputs. We apply Granger causality test respectively on two groups: online emotion time series and stock market time series. Delaying time is set to 1, 2, 3, 4 and 5 days. According to different delaying time, we calculate the $p$-value to determine the results of hypothesis test. Here, significance level is set to 5\%.

\begin{table*}[!t]
\centering
\caption{Results of Granger causality test of online emotion and stock market time series. Only significant results are listed because of the limited space. $p$-value$<0.05$: *, $p$-value$<0.01$: **, $p$-value$<0.001$.}
\label{tab:granger_causality_test}
\begin{tabular}{c|c|lllll}
\toprule
\textbf{emotion}          & \textbf{lag (days)} & \textbf{Close} & \textbf{Open} & \textbf{High} & \textbf{Low} & \textbf{Volume} \\ \hline

\multirow{5}{*}{anger}    & 1             &                &               &               &              &                 \\
                          & 2             &                &               &               &              &                 \\
                          & 3             &                &               &               &              &                 \\
                          & 4             &                &               &               &              &                 \\
                          & 5             &                &               &               &              &                 \\ \hline
\multirow{5}{*}{disgust}  & 1 & $0.0057^{**}$ & & & $0.0322^{*}$ & \\ 
                          & 2 & $0.0062^{**}$ & & & & \\
                          & 3 & & $0.0067^{**}$ & & & \\
                          & 4 & & $0.0190^{*}$ & & & \\
                          & 5 & & & $0.0280^{*}$ & & \\ \hline
\multirow{5}{*}{joy}      & 1 & & $0.0005^{***}$ & $0.0234^{*}$ & & \\ 
                          & 2 & & $6.e-5^{***}$ & $0.0304^{*}$ & & \\
                          & 3 & & $2.e-5^{***}$ & $0.0087^{**}$ & $0.0058^{**}$ & \\
                          & 4 & & $7.e-5^{***}$ & $0.0385^{*}$ & $0.0352^{*}$& \\
                          & 5 & & $0.0006^{***}$ & & & \\ \hline
\multirow{5}{*}{sadness}  & 1 & & & $0.0115^{*}$ & $0.0272^{*}$ & \\
                          & 2 & & & $0.0224^{*}$ & & \\
                          & 3 & & & $0.0303^{*}$ & & \\
                          & 4 & & & & & \\
                          & 5 & & & & & \\ \hline
\multirow{5}{*}{fear}     & 1 & & $0.0001^{***}$ & & & \\ 
                          & 2 & & $2.e-5^{***}$ & & & \\
                          & 3 & & $3.e-6^{***}$ & & & \\
                          & 4 & & $6.e-6^{***}$ & & & \\
                          & 5 & & $0.0002^{***}$ & & & \\
\hline
\end{tabular}
\end{table*}

We list testing results whose $p$-values are required to different significant levels in Table~\ref{tab:granger_causality_test}. According to the results of Granger causality test, the null hypothesis, $X_{disgust} (lag=1,2)$ series do not predict $Y_{close}$, with a high level of confidence ($p$-value$<0.01$) can be rejected. However, the other emotions do not have causal relations with $Y_{close}$. $Y_{open}$ and $X_{joy}$ ($p$-value$<0.001$), $X_{fear}$ ($p$-value$<0.001$) and $X_{disgust}$  ($p$-value$<0.05$ or even $0.01$) have causal relations. $X_{joy}$, $X_{sadness}$ and $X_{disgust}$ have causal relations with $Y_{high}$ and $Y_{low}$ ($p$-value$<0.05$ or even $0.01$). At last, results indicates trading volume in stock market time series do not have significant causal relation with any emotion time series ($p$-value$\geq0.05$). It's surprising to find that $X_{anger}$ in online emotion time series does not have causal relation with any attribute of stock market in China.

The above analysis shows that $X_{disgust}$, $X_{joy}$, $X_{sadness}$ and $X_{fear}$ can be promising features for the stock prediction models, except for $X_{anger}$.

\section{Predict the stock market}
Firstly, in this section, based on discretization methods, regression problems of predicting the stock market are converted to corresponding classification problems. Next, we perform linear and non-linear methods to solve the classification problems of stock market prediction. Eventually, the classification models are validated by 5-fold cross-validation on train set and we obtain a group of high-performance prediction models named SVM-ES.

For the prediction issue, we make use of the online emotion time series set (composed by shifted time series with different lags ranging from 1 to 5 for five emotions) or its subsets within the period from December 1st 2014 to September 9th 2015. Setting the longest lag to 5 trading days, the actual stock market time series are $Y$ from December 8th 2014 to September 16th 2015.

\subsection{Discretization of stock market data}
As illustrated in the previous sections, $Y_{close}, Y_{open}, Y_{high}$, $Y_{low}$ and $Y_{volume}$ are our targets of prediction in the stock market. Investors always just care for whether $Y_{close,i}$ (the item on $i$-th day in $Y_{close}$) are positive or negative, which will help investors make decisions to conduct stock transactions, and the binary classification (positive or negative) of $Y_{close}$ and $Y_{open}$ are also the part of our targets for prediction.

Besides, we convert regression problems of predicting five attributes in the stock market to classification problems by discretization methods that we classify each of attributes to three categories. Specifically, $Y_{close}, Y_{open}, Y_{high}$, and $Y_{low}$ are divided into three categories: \textbf{bearish}(-1), \textbf{stable}(0) and \textbf{bullish}(1) represented by CLOSE, OPEN, HIGH and LOW below. $Y_{volume}$ are divide into three categories: \textbf{low}(-1), \textbf{normal}(0) and \textbf{high}(1) represented by VOLUME below.

\begin{figure*}[!t]
\centering
\subfloat[]{\includegraphics[width=0.32\textwidth]{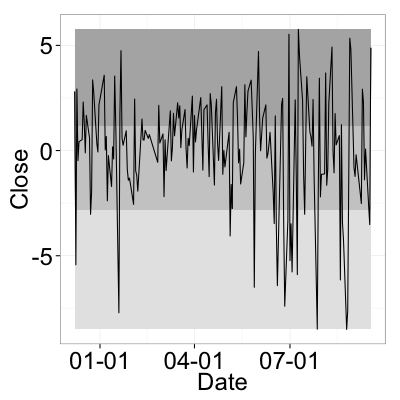}
\label{}}
\hfil
\subfloat[]{\includegraphics[width=0.32\textwidth]{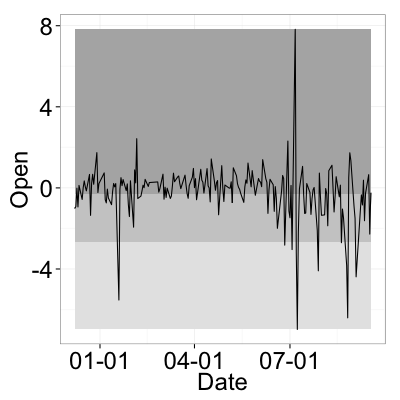}
\label{}}
\hfil
\subfloat[]{\includegraphics[width=0.32\textwidth]{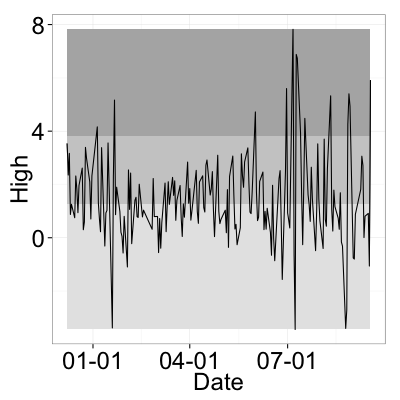}
\label{}}
\hfil
\subfloat[]{\includegraphics[width=0.32\textwidth]{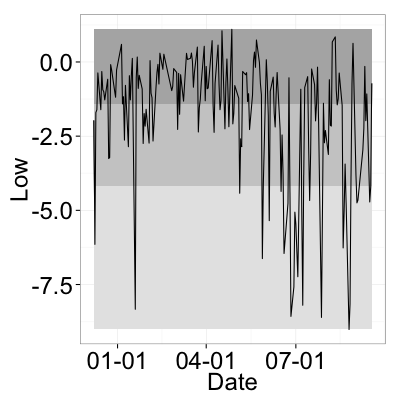}
\label{}}
\hfil
\subfloat[]{\includegraphics[width=0.35\textwidth]{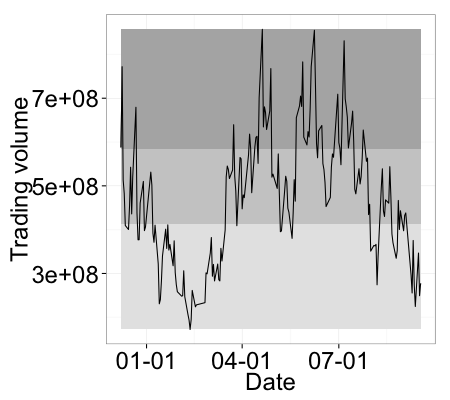}
\label{}}
\caption{Stock market time series and discretization results (by $K$-means) of $Y_{close}, Y_{open}, Y_{high}$, $Y_{low}$ and $Y_{volume}$.}
\label{fig:discretiztion}
\end{figure*}

The discretization of five attributes in the stock market is conducted by two methods: equal frequency and $K$-means clustering. Equal frequency discretization is a simple but effective method that we sort items from large to small then cut them into 3 clusters of even size. $K$-means clustering, another method we use, is popular for cluster analysis in data mining. In this paper, $K$-means clustering aims to partition observations of the stock market into 3 clusters in which each observation belongs to the cluster with the nearest distance. The results of three categories discretization by $K$-means are shown in Figure~\ref{fig:discretiztion} with 3 different grey levels.

Intuitively, as compared to the approach of equal frequency, the discretization based on $K$-means is more flexible and adjustable to the dynamic of the market. The categories it generates can better reflect the actual market status and thus can offer us a better benchmark to test the prediction results.

\subsection{Classification model for stock prediction}
In this paper, we perform machine learning methods: Logistic Regression (linear) and Support Vector Machine (non-linear) to solve the classification problems for stock prediction. These methods are both popular for training binary or multiple classification. To predict the categories $(-1, 0, 1)$ or $(0, 1)$ (just for CLOSE and OPEN) of $Y$ on $i$-th day, the input attributes of our Logistic Regression model (LR) and Support Vector Machine model (SVM) include only online emotion values of the past 5 days or a subset of them, except for other variables in the field of finance. We adapt 5-fold cross-validation to examine the accuracies of models.

At the outset, we consider all five emotions of the past 5 days as the input attributes of LR and SVM. The accuracies of models by 5-fold cross-validation are shown in Table~\ref{tab:cv_result} (3-categories and 2-categories). For the classification problem in this paper, the performance of SVM is always better than that of LR. Therefore, we infer that, relation between online emotions and the stock market is not simply linear, and the relation is more likely complicated and nonlinear.

While 3-categories discretization $(-1, 0, 1)$ results in the stock market as predicted targets, $K$-means clustering is always better than equal frequency discretization. In other words, the accuracies of models by 5-fold cross-validation, of which predicted targets are the results by $K$-means clustering, are relatively higher. Considering the categories generated by $K$-means discretization better represent the market status, we can conclude that our models indeed capture the essence of the stock fluctuation.

\begin{table}

\setlength{\abovecaptionskip}{2pt} 
\setlength{\belowcaptionskip}{2pt}
\caption{Accuracies of 5-fold cross-validation for 3-categories and 2-categories prediction models.}
\centering
\begin{tabular}{ccccccc}
\toprule
\multirow{2}*{Target (3)} & 
\multicolumn{2}{c}{equal frequency} & 
\multicolumn{3}{c}{$K$-means} \\
\cmidrule(lr){2-3}\cmidrule(lr){4-6}
& LR & SVM & LR & \textbf{SVM} & \textbf{SVM-ES}\\
\midrule
CLOSE & 34.0\%  & 43.5\%  & 52.9\%  & \textbf{58.1}\%  & 57.6\% \\
OPEN  & 37.7\%  & 44.0\%  & 53.4\%  & 61.3\%  & \textbf{64.4}\% \\
HIGH  & 36.7\%  & 39.3\%  & 48.7\%  & 53.4\%  & \textbf{54.5}\% \\
LOW   & 42.4\%  & 49.2\%  & 57.0\%  & 63.4\%  & \textbf{64.4}\% \\
VOLUME& 50.8\%  & 63.9\%  & 53.4\%  & \textbf{67.0}\%  & 66.5\% \\
\bottomrule
\label{tab:cv_result}
\end{tabular}

\begin{tabular}{ccccc}
\toprule
Target (2) & LR & SVM & SVM-ES \\
\midrule
CLOSE & 58.1\% & \textbf{61.3}\% & 60.2\% \\
OPEN  & 58.1\% & \textbf{66.0}\% & 64.9\% \\
\bottomrule
\label{tab:two_classes_result}
\end{tabular}
\end{table}

However, recalling the correlation analysis and Granger causality test of online emotions and the stock market, not all the emotions play roles on predicting the stock market and the analysis results should be used for the feature selection. Consequently, we build support vector machine model based emotions selected (SVM-ES) for stock prediction (discretized by $K$-means). The input attributes are based on analysis results of Granger causality test and Pearson correlation. We select $X_{disgust}$ (ahead of 1, 2 days) for the SVM-ES to predict CLOSE, $X_{fear}$ (ahead of 1-5 days), $X_{joy}$ (ahead of 1-5 day) and $X_{disgust}$ (ahead of 3 and 4 days) as the input attributes for predicting OPEN, $X_{joy}$ (ahead of 1-4 days), $X_{sadness}$ (ahead of 1-3 day) and $X_{disgust}$ (ahead of 5 days) as the input attributes for predicting HIGH, and $X_{sadness}$ (ahead of 1 day), $X_{joy}$ (ahead of 1-3 day) and $X_{disgust}$ (ahead of 5 days) as the input attributes for predicting LOW. Correlation analysis of $Y_{volume}$ indicates that $Y_{volume}$ and $X_{sadness}$ (ahead of 1-5 days) have the strongest correlation ($\rho>0.5$) among all online emotions, however, just using sadness as the learning feature surprisingly can not guarantee the expected performance. Thus, we try to select $X_{sadness}$ (ahead 1-5 days) and $X_{fear}$ (ahead 1-5 days) which is the second strongest relation with $Y_{volume}$ as the input attributes to predict VOLUME.

After adjusting and fixing the input attributes of SVM-ES, we train the models for stock prediction. The last column of Table~\ref{tab:cv_result} shows the accuracy of 5-fold cross-validation, respectively for 3-categories and 2-categories classification models. There are slight differences in performance between SVM-ES and the SVM trained using all the emotions, indicating emotions selected are playing dominant roles in forecasting the market. It is noteworthy that input attributes of all the SVM-ES don't include anger and it's surprising that anger shown in online social media possesses the weakest correlation or even no relation with the Chinese stock market.

From Table~\ref{tab:cv_result} it should be also noted that, emotions selected can boost the classification results attributes like OPEN, HIGH and LOW, while for CLOSE and VOLUME, SVM with all emotions as features is still the most competent solution, with slight increment (around 1\%) to SVM-ES (few attributes of input). This result explains that emotions except for input attributes of SVM-ES have very weak effects on the stock market prediction.

\subsection{Evaluation in realistic application}
\begin{figure}
\centering
 \includegraphics[width=10cm]{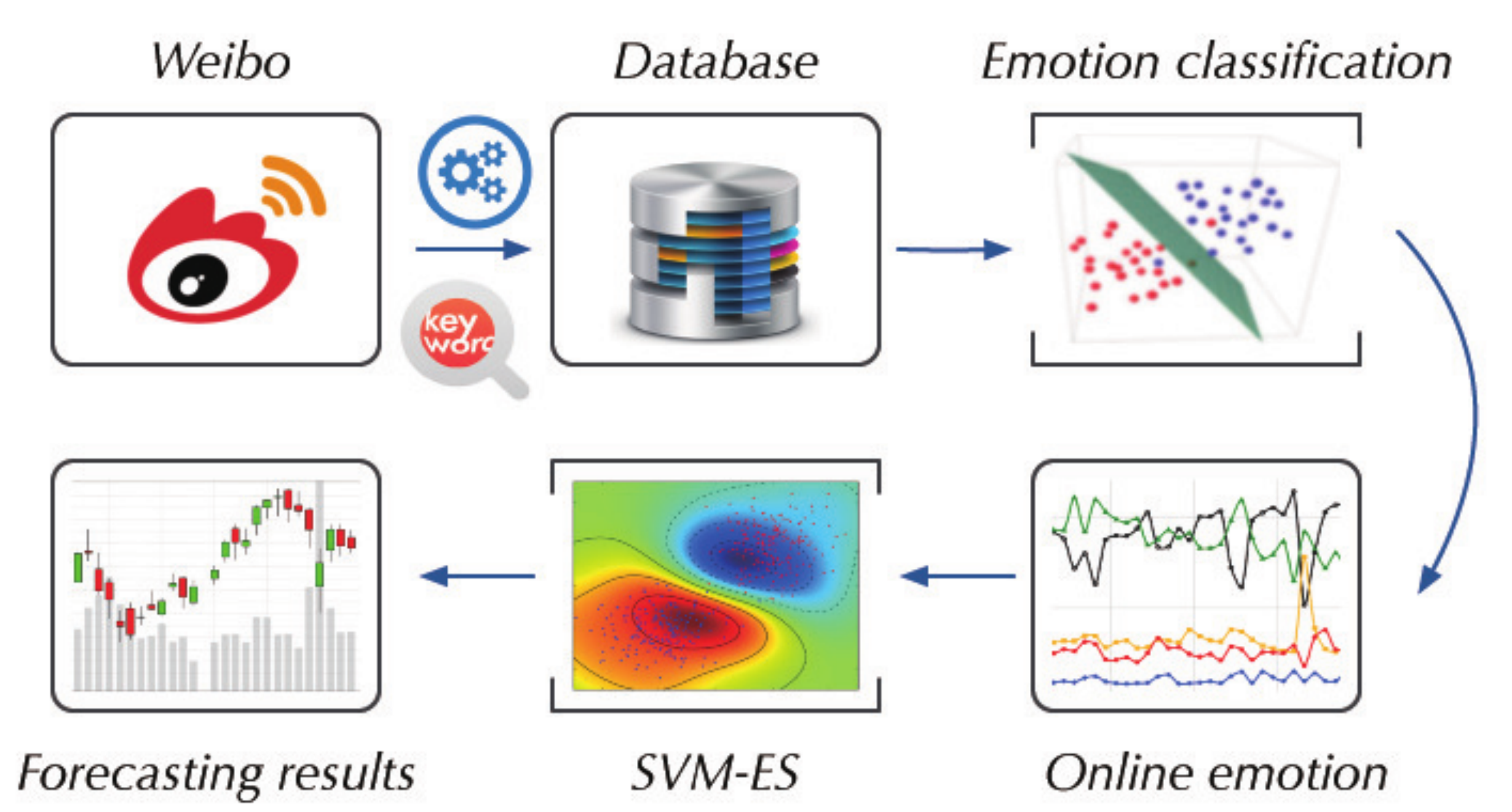}
\caption{Framework of realistic application for stock prediction based on SVM-ES.}
\label{fig:system}
\end{figure}

For further evaluating our prediction models , we sustain collecting stock-relevant tweets on Weibo with APIs and process them so as to obtain online emotion time series as our test set from September 17th to December 7th in 2015. Then we apply our classification models SVM-ES for stock prediction in the realistic Chinese stock market and we can get the daily predictions of five attributes before the market open. Framework of realistic application based on SVM-ES is depicted as Figure~\ref{fig:system}. We evaluate the stock market prediction application and the accuracies are shown in Table~\ref{tab:test_SVMES}. It turns out that the model achieves the high prediction performance, especially with accuracy of 64.15\% for the intra-day highest index (3-categories) and the accuracy of 60.38\% for the trading volume (3-categories).

\begin{table}
\caption{Accuracies of SVM-ES on realistic application.}
\centering
\begin{tabular}{ccccccc}
\toprule
CLOSE (3) & OPEN (3) & HIGH(3) & LOW(3) & VOLUME(3) & CLOSE (2) & OPEN (2) \\
\midrule
56.60\% & 43.40\% & 64.15\% & 56.60\% & 60.38\% & 60.38\% & 56.60\% \\
\bottomrule
\label{tab:test_SVMES}
\end{tabular}
\end{table}

\section{Conclusion}
In this paper, we collect massive tweets in Weibo with five categories of sentiments and focus on the stock market in China. The correlation analysis and Granger causality test are performed, which suggest that several emotions can be directly used to predict the market. Based on this, we establish several models to predict the closing index, the opening index, the intra-day highest index, the intra-day lowest index and trading volume. The results show that our model SVM-ES can outperform baseline solutions. Finally, we also testify its performance in the realistic application. In conclusion, our findings in this paper confirm that the stock market in China can be predicted by various online emotions including disgust, joy, sadness and fear.

This study has inevitable limitations, which might be interesting directions in the future work. For example, the detailed connection between the emotion and the market still remains unclear and how it evolves with time is also not discussed, however, which could help to design incremental learning schemes.

\subsubsection*{Acknowledgments.} This work was supported by the National Natural Science Foundation of China (Grant Nos. 71501005 and 71531001) and the fund of the State Key Lab of Software Development Environment (Grant Nos. SKLSDE-2015ZX-05 and SKLSDE-2015ZX-28).


\end{document}